\documentclass[conference]{IEEEtran}
\IEEEoverridecommandlockouts
\usepackage{cite}
\usepackage{amsmath,amssymb,amsfonts}
\usepackage{algorithmic}
\usepackage{graphicx}
\usepackage{textcomp}
\usepackage{xcolor}
\usepackage{amsthm}
\usepackage{color}
\usepackage{array}
\usepackage{subfig}
\usepackage{url}
\usepackage{hyperref}

\usepackage{enumitem}
\usepackage{amsfonts,bm,mathtools}
\usepackage{bbold}
\usepackage{multirow}
\usepackage[ruled,linesnumbered,noresetcount,vlined]{algorithm2e}
\usepackage{booktabs}

\newcommand{\cO}{\mathcal{O}}
\newcommand{\bx}{\bm{x}}

\newcommand{\RR}{\mathbb{R}}

\usepackage{float} 
\floatstyle{ruled}
\newfloat{model}{thp}{lop}
\floatname{model}{Model}

\def\BibTeX{{\rm B\kern-.05em{\sc i\kern-.025em b}\kern-.08em
    T\kern-.1667em\lower.7ex\hbox{E}\kern-.125emX}}
\begin{document}

\title{Load Encoding for Learning AC-OPF\\
\thanks{
This research is partly funded by both the 
NSF Awards 2007095, 2007164, 2143706, 2112533 (AI4OPT), and the
ARPA-E Perform Award AR0001136.
}
}

\author{
\IEEEauthorblockN{Terrence W.K. Mak\IEEEauthorrefmark{1}, Ferdinando Fioretto\IEEEauthorrefmark{2},
 and Pascal Van~Hentenryck\IEEEauthorrefmark{1}}
\IEEEauthorblockA{
\IEEEauthorrefmark{1} Georgia Institute of Technology, Atlanta, GA, USA
(wmak@gatech.edu, pvh@isye.gatech.edu)
}
\IEEEauthorblockA{
\IEEEauthorrefmark{2} Syracuse University, New York, NY, USA
(ffiorett@syr.edu)}
}

\maketitle

\begin{abstract}
The AC Optimal Power Flow (AC-OPF) problem is a core building block in
electrical transmission system.
It seeks the most economical active and reactive generation dispatch
to meet demands while satisfying transmission operational limits.
It is often solved repeatedly, especially
in regions with large penetration of wind farms to avoid
violating operational and physical limits. 
Recent work has shown that deep learning
techniques have huge potential in providing accurate approximations of
AC-OPF solutions. 
However, deep learning approaches often suffer from scalability
issues, especially when applied to real life power grids. 
This paper focuses on the scalability limitation 
and proposes a load compression embedding scheme to reduce 
training model sizes using a 3-step approach. 
The approach is
evaluated experimentally on large-scale test cases from the PGLib,
and produces an order of magnitude improvements 
in training convergence and
prediction accuracy.
\end{abstract}


\section{Introduction}
The \emph{AC Optimal Power Flow} (AC-OPF) is an optimization model
that finds the most economical generation dispatch meeting the
consumer demand, while satisfying the physical and operational
constraints of the underlying power network~\cite{OPF}. The AC-OPF,
together with its approximations and relaxations, constitutes a
fundamental building block for day ahead and real time market operations,
including the day-ahead security-constrained unit commitment 
and real time security-constrained economic dispatch. 

The non-convexity of the OPF limits the solving frequency of many operational
tools. In practice, generation dispatch 
in real time markets are often required
to be cleared in every couple of minutes.
Additionally, the
integration of renewable energy and demand response
programs 
create significant
stochasticity in load and generation.  
To improve solving efficiency, 
recently there has been a growing interest in applying 
machine learning techniques to power system optimization problems.
One line of research focus on how to predict
AC-OPF solutions directly using Deep Neural Networks (DNN)~(e.g.,
\cite{fioretto2020predicting,yan20real,pan19deepopf,OPFLearningTutorial21}).
Once a deep neural network is trained, 
predictions can be computed in the order of
milliseconds with a single forward pass through the network.
Deep neural networks can also be spatially decomposed~\cite{chatzos21spatial} 
to learn large-scale power networks. 


Prior results
are encouraging, showing that deep learning techniques can approximate 
solutions with high quality.  However, many of these learning
models tend to have a very large number of input features and training parameters.
For deep learning, re-training Deep Neural Nets with a 
large number of features and parameters is 
computationally challenging,  
in particular when generator commitments and grid topology 
are changing due to operators/market decisions 
and DNN needs to be frequently updated within the day-ahead/realtime markets.
Avoiding re-training completely is also computationally challenging 
since crafting a generic learning framework 
on a complex engineering system considering for all possible 
generator commitments, renewable forecasts, 
topological changes, and operator decisions 
is extremely difficult and intractable. 

\begin{figure}[t]
\centering
\includegraphics[width=0.55\linewidth]{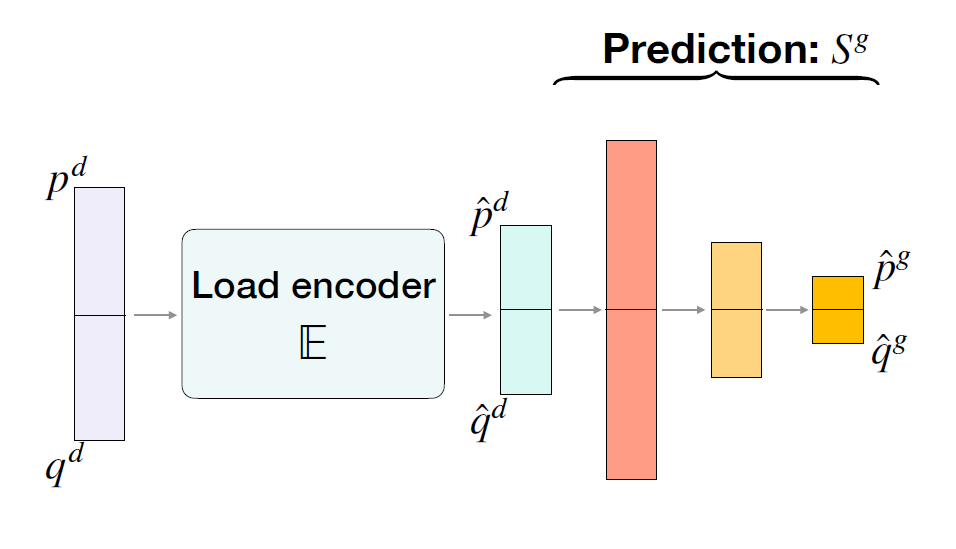}
\caption{OPF-DNN Training Architecture with Encoder $\mathbb{E}$.}
\label{fig:NN-draw}
\vspace{-7mm}
\end{figure}
This paper 
tackles the the scalability issues
of deep neural networks for learning AC-OPF 
when the network scales with increasing 
load demand input features.
Its main contribution is a 
load embedding scheme that reduces the input feature dimension 
of the deep learning network. The approach is based
on the recognition that, in many circumstances, {\em aggregating loads at
adjacent buses does not fundamentally change the nature of the AC-OPF predictions.}
The load embedding scheme has two key components:
(1) An optimization model for load aggregation that reduces the number of loads in an OPF instance, while staying close to the optimal AC-OPF cost; and 
(2) A learning model for load embedding that, given the loads for an AC-OPF instance, returns a vector of encoded loads of smaller dimensions. It is important to emphasize that the 
proposed approach is not a generic network reduction technique 
for preserving the network physical properties. 
Its goal is to reduce the input feature dimension of the learning model to 
maintain decent training accuracy within a time limit.


Figure \ref{fig:NN-draw} shows the training architecture.
The encoder first computes a load embedding, which
will then be used as inputs to the learning model to predict the
AC-OPF solutions (e.g., active \& reactive generator dispatch). 
The encoder and
AC-OPF learning models do not share parameters and are trained in
sequence --- first learns the encoder and then
trains the AC-OPF DNN using the outputs of the learned encoder. 

The approach has been evaluated on a wide range of PGLib (formerly NESTA) test cases 
\cite{Coffrin14Nesta,babaeinejadsarookolaee2021power}. 
Results show that the proposal 
can produce significant dimensionality reduction
and significant improvements in convergence speed and accuracy.

\section{Related Work}
\label{sec:related}

Power network reduction techniques, such as Kron and Ward
reduction \cite{ward49equivalent} and Principle Component
Anaylsis \cite{amchin52analyses}, have been widely used in the
industry for more than 70 years.  Early techniques focused on crafting
simpler equivalent circuits to be used by system operators for
analysis.  With the advancement of computer technology and lower
computation costs, complex reduction models became more feasible~\cite{jang13line,caliskan12kron,nikolakakos18reduced,jiang20enhanced}.
While it is possible to use classical reduction techniques to reduce
the power systems before constructing the learning model,
the learned model can only predict the reduced networks, with potential accuracy issues.
The main focus of the paper is \textit{\textbf{not}} on general network reduction techniques.
Instead, it focuses on reducing the size of the input features and 
learnable parameters 
while retaining the prediction accuracy.
 
Dimension reduction is an important and widely studied topic in
machine learning, and reduction techniques have been successfully
applied on various learning applications in power systems.  For
example, auto-encoders have been applied to predict renewable
productions~\cite{tasnim2017autoencoder,gensler16deep}, and to detect false data
injection attacks \cite{kundu20a3d}. The proposed approach differs
from general auto-encoder techniques on several aspects. First, the
load embeddings are explicitly computed through a bilevel
optimization model and not implicitly trained by an auto-encoder.
Second, the computed load embeddings are AC-feasible, and their
optimal power flows have the same cost as the original
ones.  Third, the reduced dimensionality is determined by optimization
models instead of chosen a-priori.

\section{Background}
\label{sec:background}

This paper uses the rectangular form for complex power $S = p + jq$
and line/transformer admittance $Y = g + jb$, where $p$ and $q$ denote
active and reactive powers, and $g$ and $b$ denote conductance and
susceptance.  Complex voltages are in polar form $V = v
e^{j \theta}$, with magnitude $v = \lvert V \rvert$ and phase angle
$\theta = \angle V$.  Notation $x^*$ is used to represent the complex
conjugate of quantity $x$ and notation $\hat{x}$ the prediction of
quantity $x$.

\subsection{AC Optimal Power Flow}

The AC Optimal Power Flow (OPF) determines the most economical
generation dispatch balancing the load and generation in a power
network (grid).  A power network $\bm{\mathcal N}$ is represented as a
graph $(N, E)$, where the set of nodes $N$ represent buses and the set
of edges $E$ represent branches. 
Since edges in $E$ are directed, $E^R$ is used to
denote arcs in the reverse direction.  The AC power flow equations are
expressed in terms of complex quantities for voltage $V$, admittance
$Y$, and power $S$.  Model~\ref{model:ac_opf} presents 
an AC OPF
formulation, with variables and parameters in the \textbf{complex domain}
for ease of presentation.
Superscripts $u$ and $l$ are used to indicate upper and lower bounds
for variables. The objective function ${\cO}(\bm{S^g})$ captures the
cost of the generator dispatch, with $\bm{S^g}$ denoting the vector of
generator dispatch values $(S^g_i \:|\: i \in N)$.  Constraint
\eqref{eq:ac_0} sets the voltage angle of an arbitrary slack bus $s
\in N$ to zero to eliminate numerical symmetries.  Constraints
\eqref{eq:ac_1} bound the voltage magnitudes for every bus, and constraints
\eqref{eq:ac_2} limit the voltage angle differences for every branch.
Constraints \eqref{eq:ac_3} enforce the generator output $S^g_i$ to
stay within its limits $[S^{gl}_i, S^{gu}_i]$.  Constraints
\eqref{eq:ac_4} impose the line flow limits $s_{ij}^{u}$ on all the
line flow variables $S_{ij}$.  Constraints \eqref{eq:ac_5} capture
Kirchhoff's Current Law enforcing the flow balance of generations
$S^g_i$, loads $S^d_i$, and branch flows $S_{ij}$ across every node.
Finally, constraints \eqref{eq:ac_6} capture Ohm's Law describing the
nonlinear and nonconvex AC power flow $S_{ij}$ across lines/transformers.

\begin{model}[t]
	{\small
	\caption{$\mathcal{O}({\bm{S^d}})$: AC Optimal Power Flow}
	\label{model:ac_opf}
	\vspace{-6pt}
	\begin{align}
        \mbox{\bf input:} \;\; & S^d_i\;\; \forall i\in N \nonumber \\
		\mbox{\bf variables:} \;\;
		& S^g_i, V_i \;\; \forall i\in N, \;\;
		  S_{ij} 	 \;\; \forall(i,j)\in E \cup E^R \nonumber \\
		\mbox{\bf minimize:} \;\;
		& \sum_{i \in N} {c}_{2i} (\Re(S^g_i))^2 + {c}_{1i}\Re(S^g_i) + {c}_{0i} \label{ac_obj} \\
		\mbox{\bf subject to:} \;\; 
		& \angle V_{s} = 0, \;\; s \in N \label{eq:ac_0} \\
		& {v}^l_i \leq |V_i| \leq {v}^u_i 		\;\; \forall i \in N \label{eq:ac_1} \\
		& {\theta}^{l}_{ij} \leq \angle (V_i V^*_j) \leq {\theta}^{u}_{ij} \;\; \forall (i,j) \in E  \label{eq:ac_2}  \\
		& {S}^{gl}_i \leq S^g_i \leq {S}^{gu}_i \;\; \forall i \in N \label{eq:ac_3}  \\
		& |S_{ij}| \leq {s}^u_{ij} 					\;\; \forall (i,j) \in E \cup E^R \label{eq:ac_4}  \\
		& S^g_i - {S}^d_i = \textstyle\sum_{(i,j)\in E \cup E^R} S_{ij} \;\; \forall i\in N \label{eq:ac_5}  \\ 
		& S_{ij} = {Y}^*_{ij} |V_i|^2 - {Y}^*_{ij} V_i V^*_j 			 \;\; \forall (i,j)\in E \cup E^R \label{eq:ac_6}
	\end{align}
	}
	\vspace{-15pt}
\end{model}

\subsection{Deep Learning Models}
Deep Neural Networks (DNNs) are learning architectures composed of a
sequence of layers, each typically taking as inputs the results of the
previous layer. 
Feed-forward neural networks are
DNNs where the layers are fully connected and the function
connecting the layers is given by
$
\bm{y} = \pi(\bm{W} \bm{x} + \bm{b}),
$ where $\bx \in \RR^n$ is an input vector with dimension $n$,
$\bm{y} \in \RR^m$ is the output vector with dimension $m$, $\bm{W}
\in \RR^{m \times n}$ is a matrix of weights, and $\bm{b} \in \RR^m$
is a bias vector. Together, $\bm{W}$ and $\bm{b}$ define the trainable
parameters of the network. The activation function $\pi$ is usually non-linear
(e.g., a rectified linear unit (ReLU)). This paper uses the following OPF-DNN models from
~\cite{fioretto2020predicting} to evaluate the proposed
embedding scheme:
\begin{align}
 \bm{h}_1 &= \pi(\bm{W}_1 \bm{x} + \bm{b}_1), \;\; \bm{h}_2 = \pi(\bm{W}_2 \bm{h}_1 + \bm{b}_2), \nonumber \\
 \bm{y}   &= \pi(\bm{W}_3 \bm{h}_2 + \bm{b}_3)  \label{generic_NN} 
\end{align}%
where the input vector $\bm{x} = (\bm{p^d}, \bm{q^d})$ represents the
vector of active and reactive loads, and the output vector $\bm{y}$
represents the vector of active and reactive generation
dispatch predictions $\bm{y} = (\bm{p^g}, \bm{q^g})$. 
Learning DNN models consists in finding matrices $\bm{W}$, 
and the associated bias vectors $\bm{b}$, 
to make the output prediction $\bm{\hat{y}}$ close to the
ground truth $\bm{y}$, as measured by a loss function $\mathbb{L}$.

\section{Dimensionality Reduction by Load Embedding}
\label{sec:encoding}

This section motivates the concept of load embedding, which is
formalized using a bilevel optimization model. 
The load embedding is motivated by the fact that real-life transmission systems are 
large and involve tens of thousands of buses and loads.
Naively incorporating all the input features of a large system would 
easily result in a humongous neural network,  
which is difficult to train and computationally challenging 
to cope with time limits in practice.
The proposed dimensionality reduction is motivated by the observation that, unless there is significant congestion or line power losses, moving a unit of load between two adjacent buses will not have a major effect on the final dispatch.  
This observation yields opportunities to aggregate load features into smaller subsets and reduce the 
number of training parameters. 
This section explores an encoder that performs such an aggregation.

\subsection{The Bilevel Load-Embedding Model $\mathcal{M}_{\text{BL}}$}

Let $\mathcal{O}^o$ be the optimal cost of the original OPF,
$S^{g,o}_i$ the original dispatch of
generator $i$, and $S^{d,o}_i = p^{d,o}_i + j q^{d,o}_i$ the
original complex load $i$.  The load-embedding model 
can be formulated as a bilevel optimization model $\mathcal{ M}_{\text{BL}}$:
\begin{align}
\min \;\;& \sum_{i \in N} \mathbb{1}{(S^d_i \neq 0)} \label{bl_0}\\
\mbox{s.t. }\;\;
    &\eqref{eq:ac_0} - \eqref{eq:ac_6} & \mbox{ (AC Power Flow)} \label{bl_5} \\
    & S^g_i = S^{g,o}_i \;\; \forall i \in N, &\mbox{(Generation Equiv.)} \label{bl_2}\\
    & \sum_{i \in N} p^d_i = \sum_{i \in N} p^{d,o}_i &\mbox{(Active Load Equiv.)}\label{bl_3} \\
    & \sum_{i \in N} q^d_i = \sum_{i \in N} q^{d,o}_i &\mbox{(Reactive Load Equiv.)} \label{bl_4} \\
    & \lvert \mathcal{O}(\bm{S^d}) - \mathcal{O}^o \rvert \leq \beta & \mbox{(Cost Equiv.)} \label{bl_1}
\end{align}
Its goal is to find the embedded loads $(p^d_i,q^d_i)$ ($i \in N$)
which are the key decision variables. Objective \eqref{bl_0} minimizes
the number of nonzero loads using an indicator function. Constraints
\eqref{bl_5} impose the power flow equations. Constraints
\eqref{bl_2} ensure that the generation dispatch remains the same,
given that they are the targets of the learning task.  Constraints
\eqref{bl_3} and \eqref{bl_4} require the sum of the active and
reactive loads to remain the same after the encoding. Together, these
constraints ensure that the loads are AC-feasible for the original
generation dispatch. However, they do not guarantee that they could
not be served by a significantly better generator dispatch. This is
the role of constraint \eqref{bl_1} that ensures the encoded load
vector $\bm{S^d} = \bm{p^d} + j \bm{q^d}$ induces an optimal flow with
cost close to the original cost $\mathcal{O}^o$ (within a tolerance
parameter $\beta$). This constraint introduces an AC-OPF as a
subproblem, hence creating a bilevel model.

\subsection{Load Embedding with Congestion Constraints: Model $\mathcal{M}_{\text{R}}$}

Optimization model $\mathcal{M}_{\text{BL}}$ is challenging for two
reasons: (1) it implicitly features discrete variables through the
indicator variables in its objective; (2) it is a bilevel optimization
problem. The first challenge can be addressed by replacing its discrete
objective by a continuous expression that maximizes the square of
the real and reactive powers of each load, i.e., 
\begin{align}
\max \;\;& \sum_{i \in N} [(p^d_i)^2 + (q^d_i)^2]. \label{r_0}
\end{align}
Objective \eqref{r_0} encourages active and reactive loads to be
aggregated without the need of binary variables.

The second challenge can be addressed by replacing constraint
\eqref{bl_1} by proxy constraints that characterize the OPF.  Indeed,
in the original OPF, a number of voltage and thermal constraints are
binding. Imposing constraints on the associated voltages and flows
will help in keeping the optimal cost close to the original cost. 
Let $N^{l}$ and $N^{u}$ be the set of buses with binding lower and upper
constraints on voltages, and $E^{u}$ be the set of lines with binding
thermal limit constraints. Constraint \eqref{bl_1} can be
relaxed and reformulated as:
\begin{align}
    \lvert v_i - v^u_i \rvert \leq \beta_v, \forall i \in N^u \label{r_1} &\mbox{ (Voltage Congestion)}\\
    \lvert v_i - v^l_i \rvert \leq \beta_v, \forall i \in N^l \label{r_2} &\mbox{ (Voltage Congestion)}\\
    \lvert \lvert S_{ij} \rvert - s^u_{ij} \rvert \leq \beta_s, \forall (i,j) \in E^u  &\mbox{ (Line Congestion)}\label{r_3}
\end{align}
where $\beta_v$ and $\beta_s$ are the tolerance parameters for the
tightness of the original binding constraints. The
relaxed model $\mathcal{M}_{\text{R}}$ is then defined as:
\begin{align}
\max \;\;& \sum_{i \in N} [(p^d_i)^2 + (q^d_i)^2] \nonumber \\
\mbox{s.t. }\;\;  & \eqref{eq:ac_0} - \eqref{eq:ac_6} & \mbox{ (AC Power Flow)} \nonumber\\
                      & \eqref{bl_2} - \eqref{bl_4}& \mbox{(Equiv. Constr.)} \nonumber \\
 	              & \eqref{r_1} - \eqref{r_3} & \mbox{(Congestion Constr.)} \nonumber
\end{align}

\subsection{Load Embedding with a Penalty Method: Model $\mathcal{M}_{\text{P}}$}

Model $\mathcal{M}_{\text{R}}$ requires the choice of tolerance
parameters $\beta_v$ and $\beta_s$.  If these tolerances are too
tight, it may not be possible to aggregate loads effectively. If they
are too loose, the resulting predictions may be inaccurate.  To
overcome this difficulty, this paper uses a penalty
method.\footnote{Alternatively, it is possible to use an Augmented
  Lagrangian Method. Experimental results have shown that the encoding
  quality is similar but solving times were slightly longer for the
  latter.}  The resulting model $\mathcal{M}_{\text{P}}(\beta_v,
\beta_s)$ becomes
\begin{align}
\max \;\;& \sum_{i \in N} [(p^d_i)^2 + (q^d_i)^2] +  
\beta_{v} \sum_{i \in N^u}  \lVert v_i - v^u_i \rVert^2 +  \nonumber \\
&\beta_{v} \sum_{i \in N^l} \lVert v_i - v^l_i \rVert^2 + 
\beta_{s} \sum_{(i,j) \in E^u} \lVert \lvert S_{ij} \rvert - s^u_{ij} \rVert^2 \nonumber\\
\mbox{s.t.} \;\;&  \eqref{eq:ac_0} - \eqref{eq:ac_6} \mbox{ and } \eqref{bl_2} - \eqref{bl_4} \nonumber
\end{align}
and it can be solved iteratively by increasing $\beta_v$ and $\beta_s$
until \eqref{bl_1} is satisfied, using Algorithm~\ref{alg:penalty}.


\begin{algorithm}[t]
\small
  \caption{Load Encoding}
  \label{alg:penalty}
  \DontPrintSemicolon
  \setcounter{AlgoLine}{0}
  \SetKwInOut{Input}{Input}
  \SetKwInOut{Output}{Output}
  \Input{$\mathcal{N}:$ power grid data; 
                 $\rho_v, \rho_s:$ penalty steps;\\
                 $(\beta_v, \beta_s):$ constraint tolerances; \\
                 $\beta: $ cost tolerance;\\
  		 $i^u:$ max iteration limit.}
  \Output{$\bm{S^d} = (\bm{p^d}, \bm{q^d})$}
  \label{line:1}
  \For{$i = 0, 1, 2, \ldots, i^{u}$} { 
    $ \bm{S^d} \gets (\bm{p^d}, \bm{q^d}) \gets \mathcal{M}_{\text{R}}(\beta_v, \beta_s)$\\
    \If{$\lvert \mathcal{O}(\bm{S^d}) - \mathcal{O}^o \rvert \leq \beta $}  {\textbf{break}}
    $\beta_v \gets \rho_v \beta_v, \;\; \beta_s \gets \rho_s \beta_s$ \\
   }
\end{algorithm}

\section{Learning to Encode}
\label{sec:learningToEncode}

Algorithm \ref{alg:penalty} computes an ``optimal'' load embedding for a
load profile $\bm{S^d}$. However, computing Algorithm \ref{alg:penalty} 
at prediction time is expensive, hence defeating the purpose of speed up OPF computations. 
Instead, this paper proposes {\em to learn the encoder}, i.e., to learn a machine-learning
proxy for Algorithm \ref{alg:penalty}. The idea is to take the set of training
instances for OPF and apply Algorithm \ref{alg:penalty} to obtain the load
embeddings. The machine-learning model then learns the mapping between the
original and embedded loads. The input vector is the load vector 
$\bm{x} = (\bm{p^d}, \bm{q^d})$ and the output vector is the embedded load vector $\bm{z} =
(\bm{\hat{p}^{d}}, \bm{\hat{q}^{d}})$. The structure of the output,
i.e., the embedded load vector, is obtained by removing the loads that
are {\em relocated in all the training instances}. The training of the machine-learning
model then consists in mapping the original loads into this smaller set of loads
to mimic Algorithm \ref{alg:penalty}. The paper proposes to {\em learn the encoder}, 
and explores two learning schemes: (1) a linear regression
$
\mathbb{E}_L(\bm{x}) = \bm{z} = \bm{W} \bm{x} + \bm{b},
$
that is extremely fast, and; (2) a DNN similar to \eqref{generic_NN}
$
\mathbb{E}_F(\bm{x}) =  \bm{z}   = \pi\{\bm{W}_3 \pi[\bm{W}_2 \pi(\bm{W}_1 \bm{x} + \bm{b}_1) + \bm{b}_2] + \bm{b}_3\}.
$

For the full NN-encoder, the dimension of the first and second layers are set to
twice of the dimensions of the input and the output vectors
respectively. For a data set collection $\mathbb{D} = \{ (\bm{x}^i,
\bm{z}^i) : i \in [1 \dotsc n]\}$ with $n$ test cases where the
outputs $\bm{z}^i$ are computed using Algorithm~\ref{alg:penalty}, the
goal of the learning task is to find the model parameters $\bm{W}$ and
$\bm{b}$ that minimize the empirical
risk function:
\begin{align}
    \min_{\bm{W}, \bm{b}} & \sum_{(\bm{x}^i, \bm{z}^i) \in \mathbb{D}} 
    \mathbb{L}(\mathbb{E}_{m}(\bm{x}^i), \bm{z}^i ). \label{alg:enc_train}
\end{align}
\noindent
where $m \in \{L, F\}$ is used to discriminate the model adopted. 


\section{Scalable OPF Learning}
\label{sec:scalableOPFLearning}

Once the load encoder has been learned, the OPF learning task is
performed using the architecture in Figure \ref{fig:NN-draw}.  NN
layers (tensors) are represented by rectangular boxes and arrows
represent connections between layers. The architecture uses fully
connected layers with ReLU as the activation functions. Notice that
the encoder is pre-trained, so the learning task will not affect its
parameters.
The dimensions of $h_1$ and $h_2$ of the OPF layers are
set to twice of the dimension of the input and the output vectors
respectively as \cite{fioretto2020predicting}.
\footnote{When the models for OPF and the encoder are both fully trained, only 1 forward NN pass (encoder, then OPF model) is needed during predictions.}



Unless required to, the encoder does not preserve the values of many
physical parameters, including phase angles, voltage magnitudes, and
line flows. However, these physical values on the reduced network,
which are available as a result of Algorithm~\ref{alg:penalty}, are
still important to improve prediction accuracy using, for instance, a
Lagrangian dual approach as in~\cite{fioretto2020predicting}.

\section{Experimental Evaluation}
\label{sec:evaluation}

\noindent\textbf{Parameter Setup:}
The experiments were performed on various PGLib (formerly NESTA) 
test cases \cite{Coffrin14Nesta}, and Algorithm~\ref{alg:penalty} was implemented on top of
PowerModels.jl \cite{Coffrin:18}, a state-of-the-art open source
Julia package for solving or approximating AC-OPF. The
tolerance $\beta$ was set to 0.5\%, $i^u = 500$, and parameters
$\rho_v$ and $\rho_s$ were both set to 1.5. The OPF data sets were generated by varying the load profiles of each
benchmark network from 80\% to 120\% of their original (complex) load
values, with a step size of 0.02\%, giving a maximum of 2000 test
cases for every benchmark network. For each test case, to create
enough diversity, every load is perturbed with random noise from the
polar Laplace distribution whose parameter $\lambda$ is set to 10\% of
the apparent power. 
The test cases were split with 80\%-20\% ratio
for training and testing. 
The OPF-DNN models and the encoding models
($\mathbb{E}_L/\mathbb{E}_F$), were implemented using PyTorch
\cite{paszke:17} and run with Python 3.6,
with the Mean Squared Error (MSE) as the loss function.  The training was performed using
Tesla-V100 GPUs with 16GBs HBM2 ram on machines with Intel CPU cores
at 2.1GHz. The training used Averaged Stochastic Gradient Descent (ASGD)
with learning rate $0.001$. 
The paper uses default parameters from PyTorch for 
all other hyper-parameters across various learning models. 


\begin{table}[t]
\centering
\small
\caption{OPF-DNN: Original \& Reduced Input dimension}
\resizebox{0.65\linewidth}{!}
  {
  \begin{tabular}{l|cc|c}
  \toprule
     Network & Orig. dim. & Reduced dim. & Reduction \% \\
     \midrule
     \textbf{14\_ieee}      &  22   &  11   & 50\% \\
     \textbf{30\_ieee}      &  42   &  13   & 69\% \\
     \textbf{39\_epri}      &  42   &  29   & 31\% \\
     \textbf{57\_ieee}      &  84   &  26   & 69\% \\
	\textbf{73\_ieee\_rts}  &  102  &  88   & 14\% \\
   \textbf{89\_pegase}      &  70   &  55   & 21\% \\
    \textbf{118\_ieee}      &  198  &  198  & 0\% \\
	\textbf{162\_ieee\_dtc} &  226  &  143  & 37\% \\
    \textbf{189\_edin}      & 1244 & 336 &  73\%\\
    \textbf{1394\_sop\_eir} & 524  &207  & 60\% \\
    \textbf{1460\_wp\_eir} & 536  & 536  & 0\% \\
    \textbf{1888\_rte} & 2000 & 1555 & 22\% \\
    \textbf{2848\_rte} & 3022 & 2310 & 24\% \\
    \textbf{2868\_rte} & 3102 & 2221 & 28\% \\
    \textbf{3012wp\_mp} & 4542  & 2043 & 55\% \\
    \textbf{3375wp\_mp} & 4868  & 2109 & 57\% \\
   \bottomrule
  \end{tabular}
  }
  \label{tbl:data_dim} 
  \vspace{-6mm}
\end{table} 

\subsection{Compression Ratios}
\label{sec:exp1}

Table~\ref{tbl:data_dim} shows, for each test case, the dimension
(i.e., $c_p + c_q$) for the input load tensor ($\bm{p^d}$,$\bm{q^d}$) in the
original data set and compares it to the reduced dimension for the
input load tensor ($\bm{\hat{p}^d}$,$\bm{\hat{q}^d}$). This reduced
dimension is computed by running Algorithm~\ref{alg:penalty}
on the ($\approx$ 2000) test cases for each benchmark data set and
removing the loads that are always assigned to zero. Remarkably, given
the wide range of considered loads, many of the test cases (except the
118 and 1460 cases) achieve a significant dimensionality reduction. Of
particular interest are the large RTE test cases whose dimensions are
reduced by 24\% and 28\% and the large {\tt wp\_mp} test cases whose
dimensions are reduced by 55\% and 57\%.



\begin{table}[h]
\centering
\small
\caption{Prediction Errors (p.u.).}
\resizebox{0.60\linewidth}{!}
  {
  \begin{tabular}{l|lll}
  \multicolumn{4}{c}{L1 Error for Generator Dispatch}\\
  \toprule
     Network & No Enc. & Linear Enc. & Full Enc. \\
     \midrule
     \textbf{14\_ieee}      &0.0065 &0.0057 & \bf{0.0050}\\
     \textbf{30\_ieee}      &0.0041 & \bf{0.0033} & 0.0038\\
     \textbf{39\_epri}      &0.2536 &0.0632 &\bf{0.0422}\\
     \textbf{57\_ieee}      &0.0433 &0.0522 &\bf{0.0130} \\
	\textbf{73\_ieee\_rts}  &0.0602 &\bf{0.0178} &0.0676 \\
   \textbf{89\_pegase}      &0.1807 &\bf{0.0243} &0.0360\\
    \textbf{118\_ieee}      &0.0504 &0.0108 &\bf{0.0063} \\
	\textbf{162\_ieee\_dtc} &0.1622 &0.0493 &\bf{0.0329} \\
    \textbf{189\_edin}      &0.0209 &0.0117 & \bf{0.0075}\\
    \textbf{1394\_sop\_eir} &0.0041 &0.0039 & \bf{0.0029}\\
    \textbf{1460\_wp\_eir}  &0.0129 &0.0114 & \bf{0.0055}\\
    \textbf{1888\_rte}      &0.1964 &\bf{0.0792} &0.2046 \\
    \textbf{2848\_rte}     &0.0376 & 0.0125& \bf{0.0085}\\
    \textbf{2868\_rte}     &0.025 &\bf{0.0095} & 0.2026\\
    \textbf{3375wp\_mp}     &0.0483 &0.0252 & \bf{0.0212}\\
   \bottomrule
  \end{tabular}
  }
  \label{tbl:val_disp_MSE} 
  \vspace{-5mm}
\end{table}

\subsection{OPF Prediction Errors}
\label{sec:exp3}

This section shows that learning with encoders almost always reduces
prediction errors, which is interesting in its own
right. Table~\ref{tbl:val_disp_MSE} depicts the
prediction results for three variants of OPF-DNN models on the testing
data set: a) no encoder, b) the OPF-DNN architecture with encoder
$\mathbb{E}_L$, and c) the OPF-DNN architecture with encoder
$\mathbb{E}_F$.  Table~\ref{tbl:val_disp_MSE} reports the averaged
L1-losses $\lVert \cdot \rVert_1$ for the \emph{main predictions}:
\begin{align}
      \frac{1}{ \lvert T \rvert}\sum_{t \in T}[ 
      \frac{ \lVert \bm{\hat{p}^g_t}, \bm{p^g}_t \rVert_1 / \lvert N^G \rvert +
        \lVert \bm{\hat{q}^g}_t, \bm{q^g}_t \rVert_1 / \lvert N^G \rvert}{2} 
         ] \nonumber        
\end{align}
where $N^G$ is the set of generators and $T$ is the set of testing
data. 
The results demonstrate the effectiveness of the proposed encoders,
which yield predictors with smaller errors. Interestingly, even for
the 118 bus and 1460 bus benchmarks, which have no dimensionality
reduction, the generation dispatch errors are reduced by an
order of magnitude.
The results indicate that employing encoders
are always effective, regardless on whether its a simplified linear model or a more complex multiple layered models.

\begin{figure}[h]
\centering
\vspace{-4mm}
\includegraphics[width=0.40\linewidth]{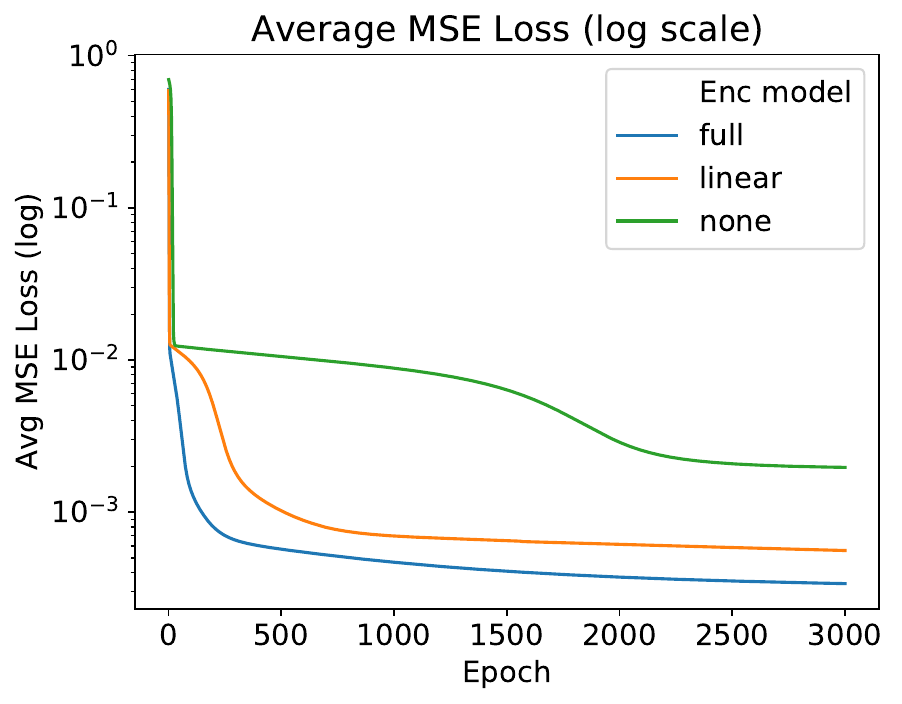}
\includegraphics[width=0.40\linewidth]{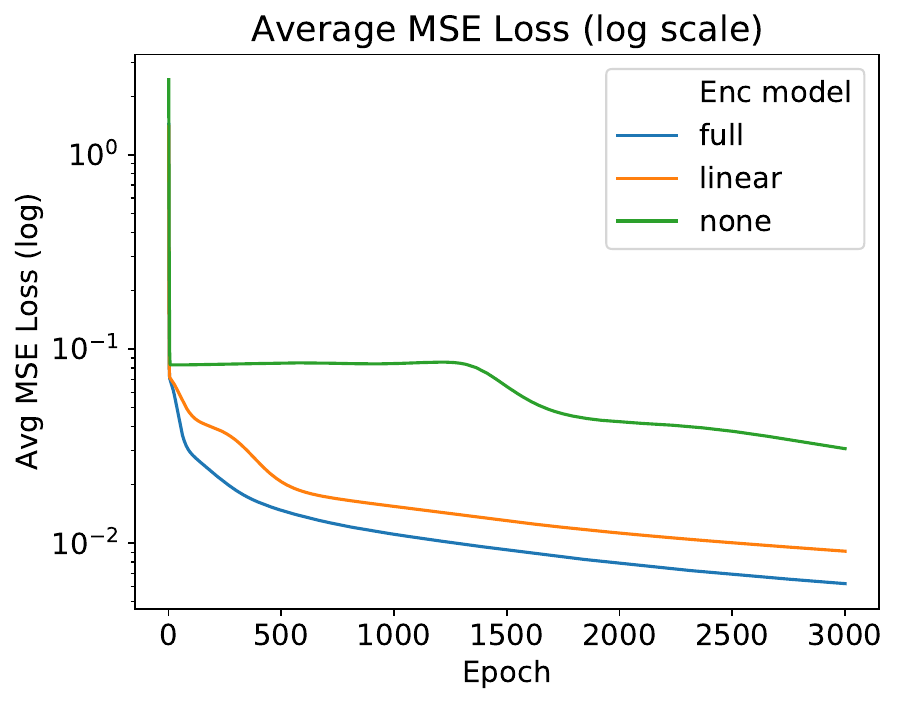}
\caption{Average MSE error (log scale) during training phase. Left: 1394\_sop\_eir, right: 3375wp\_mp.}
\label{fig:convergence}
\vspace{-5mm}
\end{figure}
\subsection{Training Convergence and Speed}
\label{sec:exp4}

The key motivation of this paper is to speed up the learning
task. This section demonstrates that the OPF-DNN architecture with
load encoding quickly converges to an accuracy that is an order of
magnitude better than full OPF-DNN architecture.
Figure~\ref{fig:convergence} shows the combined MSE losses (in log
scale) for the three OPF-DNN architectures during the training phase
for the 1394 and 3375 bus benchmarks. The results are averaged
by the number of training cases as in previous sections. The full
NN-encoder is almost an order of magnitude more accurate than the base
model, and consistently better than the linear encoder model.  Both
load-encoding architectures outperforms the base model in the early
convergence period (within 500 epochs), and a significant convergence
gap still exists even after the error curves have flattened (e.g.,
after 2500 epochs).  These results indicate that load reduction yields
both a better training convergence and smaller prediction errors.
\footnote{
Additional results indicate that training the 
linear/full encoder with good convergence can be performed in under 20 min/2 hr,
translating to roughly 20/120 extra epochs for OPF training on
the largest network. 
Clearly, Figure~\ref{fig:convergence} indicates
the benefits of encoders outweigh the extra computational burden.
}

\section{Conclusion}
\label{sec:conclusion}

This paper studied how to improve the scalability of deep neural
networks for learning the active and reactive power of generators in
AC-OPF. To address computational issues that arise in learning AC-OPF
over large networks, this paper proposed a load encoding scheme for
dimensionality reduction and its associated deep learning
architecture. The load encoding scheme consists of (1) an optimization
model to aggregate loads for each instance; and (2) a deep learning
model that approximates the load encoding. The learned encoder can
then be included in a deep learning architecture for AC-OPF and
produces an order of magnitude improvement in training convergence and
prediction accuracy of large realistic cases.

 \bibliographystyle{elsarticle-num} 
 \bibliography{dl_opf}

\end{document}